\documentclass[aps,prl,showpacs,amsmath,prd,graphicx,nofootinbib]{revtex4-1}
\usepackage{amssymb,graphics,graphicx,tabularx,color,epstopdf}
\usepackage{psfrag}

\begin{document}
\title
{\Large \bf Binary Icosahedral Flavor Symmetry\\
 for Four Generations of Quarks and Leptons}

\author{Chian-Shu~Chen$^{1,2}$\footnote{chianshu@phys.sinica.edu.tw},
Thomas W. Kephart$^{3}$\footnote{tom.kephart@gmail.com}, 
and Tzu-Chiang Yuan$^{2,4}$\footnote{tcyuan@phys.sinica.edu.tw}}

\affiliation{$^{1}$Physics Division, National Center for Theoretical Sciences, Hsinchu, Taiwan 300\\
$^{2}$Institute of Physics, Academia Sinica, Taipei, Taiwan 115\\
$^{3}$Department of Physics and Astronomy, Vanderbilt University, Nashville, Tennessee 37235, USA\\
$^{4}$Kavli Institute of Theoretical Physics China, CAS, Beijing 100190, China}
\date{\today}

\begin{abstract}
To include the quark sector, the $A_{5}\equiv I$ (icosahedron) four generation lepton model is extended to a binary icosahedral symmetry $I'$ flavor model.  We find the masses of fermions, including the heavy sectors, can be accommodated. At leading order the CKM matrix is the identity and the PMNS matrix, resulting from same set of vacua, corresponds to tribimaximal mixings. 

\noindent
\end{abstract}

\maketitle

\noindent
\section{Introduction}
 
The current version of the standard model (SM) consists of three generations 
of quarks and leptons. 
Recently we proposed \cite{Chen:2010ty} a four generation lepton model based on 
the non-abelian discrete symmetry $A_{5}\equiv I$ (icosahedron), in which the best features of 
the three family $A_{4}\equiv T$ (tetrahedral) model survive. Besides the new heavy degrees of freedom in the $A_{5}$ model, which satisfy the experimental constraints, we retain  tribimaximal neutrino mixings, three light neutrino masses, and three SM charged lepton masses in the three light generation sector.

In this paper, we will explore a generalization of our $A_5$ model to include  
four generations of both quarks and leptons. But before launching into that discussion
we must first discuss the viability of models with four generations given recent experimental developments.
A fourth generation is now being constrained \cite{Eberhardt:2010bm} by precision electroweak data \cite{NA62:2011aa}, by flavor symmetries \cite{Lacker:2010zz}, and by the  
Higgs-like particle at 125 GeV recently reported at the LHC \cite{atlas,cms,cms-search}. The new data provide an important step forward 
in distinguishing various four generation models, and in particular eliminating some from 
consideration. In particular,  four sequential generation models are now highly disfavored \cite{Eberhardt:2010bm};
however, it would be premature to dismiss all four generation models.
While tension between four generation models and data has developed, a fourth generation is not excluded by the electroweak 
precision data \cite{Holdom:2009rf}, so the existence of a fourth generation is still 
a viable phenomenological possibility which can provide an explanation of the observed anomaly 
of CP asymmetries in the B meson system \cite{Hou:2005hd}, and the baryon asymmetry 
of the universe \cite{Hou:2008xd}, with additional mixings and CP phases. 
Also, there are a number of way to relieve this tension. For example
two Higgs doublet models (See e.g., \cite{BarShalom:2011zj} and references therein) can accommodate a fourth generation of fermions and current data.
For a comprehensive review see Ref. \cite{BarShalom:2012ms}. Typically these two Higgs doublet models are low energy effective field theories that
require composite Higgses similar to top quark condensate models \cite{Bardeen:1989ds}. For some recent examples see  Ref. \cite{Hung:2009hy}.
Another possibility is to add electroweak doublets that are in color octets
\cite{Manohar:2006ga}. Further discussion can be found in Ref. \cite{He:2011ti}.
While the model we will discuss has an extended Higgs structure,  a full exploration of the possible composite nature of the scalar sector is beyond the scope of our present study.

To generalize our $A_5$ model to include  
four generations of quarks and leptons, we first recall the three family scenarios in which  
the binary tetrahedral group $T'\equiv SL_{2}(3)$ is capable of providing 
a model of both the quarks and leptons with tribimaximal  mixings and 
a calculable Cabibbo angle~\cite{Frampton:2008bz}. The $T'$ group is 
the double covering group of $A_{4}$. It  has four irreducible  representations (irreps) 
with identical multiplication rules to those of $A_{4}$, one triplet $\textbf{3}$ 
and three singlets $\textbf{1}, \textbf{1}'$, and $\textbf{1}''$, 
plus three additional doublet irreps $\textbf{2}, \textbf{2}'$, and $\textbf{2}''$. 
The additional doublets allow the implementation of the $\textbf{2}\oplus\textbf{1}$ 
structure to the quark 
sector~\cite{Aranda:1999kc,Frampton:2007et,Feruglio:2007uu,Chen:2007afa,Frampton:1994rk}, 
thus the third family of quarks are treated differently and are assigned to a singlet.
Hence they can acquire heavy masses~\cite{Frampton:1994xm,Frampton:1995wf}. 
One should note that $A_{4}$ is not a subgroup of $T'$, therefore, the inclusion of 
quarks into the model is not strictly an extension of $A_{4}$, but instead 
replaces it~\cite{Frampton:2009pr}. Based on the same philosophy, we study the model of 
four families of quarks and leptons by using the binary icosahedral group $I'\equiv SL_{2}(5)$. 
The relation between $I'$ and $A_{5}$ is similar to that for $T'$ and $A_{4}$. 
The icosahedral group $A_{5} \subset SO(3)$ has double-valued representations that 
are single-valued representations of the double icosahedral group $I' \subset SU(2)$. 
Hence, besides the  irreps of $I'$ that are coincident 
with those of $A_{5}$, there are four additional spinor-like irreps 
$\textbf{2}_{s}, \textbf{2}'_{s}, \textbf{4}_{s}$, and $\textbf{6}_{s}$ of $I'$. 
We shall be able to assign quarks to the spinor-like representations, 
but to discuss model building using $I'$, we must first review our lepton model 
based on $A_{5}$, which will remain essentially unchanged when generalized to $I'$. 
Some useful group theory details have been relegated to the Appendix.

\section{The Leptonic $A_{5}$ model}
  
The irreps of $A_{5}$ are one singlet $\textbf{1}$, two triplets $\textbf{3}$ and $\textbf{3}'$, one quartet $\textbf{4}$, and one quintet $\textbf{5}$. The model is required to be invariant under the flavor symmetry of 
$A_{5}\times Z_{2}\times Z_{3}$ and the particle content is given by Table~\ref{A5}.
\onecolumngrid
\begin{center}
\begin{table}[ht]
\begin{tabular}{|p{1.3cm}p{1.3cm}p{1.3cm}p{1.3cm}p{1.3cm}p{1.3cm}p{1.3cm}p{1.3cm}p{1.3cm}p{1.3cm}p{1.3cm}|}\hline
Field & $L_{i}$ & $l_{R5}$ & $l^c_{R3}$ & $l^{(1),(2)}_{R1}$ & $N_{R5}$ & $N^{(1)_{R}}$ & $S_{4}$ & $H_{4}$ & $H'_{4}$ & $\Phi_{3}$ \\ \hline 
$SU(2)_{L}$ & $\textbf{2}$ & $\textbf{1}$ & $\textbf{1}$ & $\textbf{1}$ & $\textbf{1}$ & $\textbf{1}$ & $\textbf{1}$ & $\textbf{2}$ & $\textbf{2}$ & $\textbf{2}$ \\
$A_{5}$ & $\textbf{4}$ & $\textbf{5}$ & $\textbf{3}$ & $\textbf{1}$ & $\textbf{5}$ & $\textbf{1}$ & $\textbf{4}$ & $\textbf{4}$ & $\textbf{4}$ & $\textbf{3}$ \\ 
$Z_{2}$ & 1 & -1 & -1 & -1 & 1 & 1 & 1 & 1 & -1 & 1 \\ 
$Z_{3}$ & $\omega$ & 1 & 1 & 1 & 1 & 1 & 1 & $\omega^2$ & $\omega^2$ & $\omega^2$ \\\hline
\end{tabular}
\caption{\label{A5} Particle content of the lepton $A_{5}$ model.}
\end{table}
\end{center} 
%
\noindent
Here $L_{i} = (\nu_{i} , l_{i})^{T}$ is the left-handed $SU(2)_{L}$ doublet with generation 
index $i$, $l_{R}$'s and $N_{R}$'s are right-handed charged leptons and neutrinos respectively. $H_{4}$, $H'_{4}$ and $\Phi_{3}$ are $SU(2)_{L}$ doublet scalar fields, while $S_{4}$ is a singlet scalar. The representations of $SU(2)_{L}$ gauge symmetry should not be confused with the representations of the non-abelian discrete symmetry $A_{5}$. The most general Yukawa interactions invariant under the symmetries can be expressed as  
%
\begin{eqnarray}
\emph{L}_{\rm {Y(lepton)}} &=& \frac{1}{2}M_{1}N^{(1)}_{R}N^{(1)}_{R} + \frac{1}{2}M_5N_{R5}N_{R5} \nonumber \\
&+& Y_{S1}(S_{4}N_{R5}N_{R5}) + Y_{S2}(S_{4}(l^{-}_{R3})^{c}l^{-}_{R5}) \nonumber \\
&+& Y_{1}(L_{L4}N^{(1)}_{R}H_{4}) + Y_{2}(L_{L4}N_{R5}H_{4})  \\
&+& Y_{3}(L_{L4}N_{R5}\Phi_{3})  + Y_{4}(L_{L4}l_{R5}H'_{4}) \nonumber \\
&+& Y_{5}(L_{L4}l_{R1}H'_{4}) + Y_{6}(L_{L4}l_{R2} H'_{4}) + \rm {H.c.} \nonumber 
\end{eqnarray}  
%
\indent
If the scalar $S_{4}$ develops the vacuum expectation value (VEV)  $\langle S_{4} \rangle = (V_{S}, 0, 0, 0)$, then $A_{5}$ will break to $A_{4}$ causing the $A_{5}$ irreps  to decompose as $\textbf{1} \rightarrow \textbf{1}$, $\textbf{3} \rightarrow \textbf{3}$, $\textbf{3}' \rightarrow \textbf{3}$, $\textbf{4} \rightarrow \textbf{1}\oplus\textbf{3}$, and $\textbf{5} \rightarrow \textbf{1}'\oplus\textbf{1}''\oplus\textbf{3}$. There is one vector-like $SU(2)_{L}$ singlet charged lepton field with the mass given by $\langle S_{4}\rangle$. The masses of the four chiral generations of charged leptons are generated by scalar field $H'_{4}$ with an interesting result that the electron mass is predicted to be zero at tree level and induced through quantum corrections. We argued in \cite{Chen:2010ty}    
that there is enough freedom to fit the observed charged lepton masses. The canonical seesaw mechanism is responsible for the left-handed neutrino masses in the $A_{5}$ model\footnote{Also see Ref.~\cite{Schmidt:2011jp} for a general discussion of neutrino masses with four generations of fermions.}, while the Dirac mass terms are provided by the scalars fields $H_{4}$ and $\Phi_{3}$. We are able to obtain one heavy neutrino and three SM light neutrino masses through arranging the VEVs of the two scalar fields without severe fine-tuning. Therefore, the mass structure of the three families under $A_{4}$ symmetry is retained. We refer the reader to Ref.~\cite{Chen:2010ty} for the details of the model. Now we discuss the inclusion of the quark sector by extending $A_{5}$ to its double covering $I'$. 

\section{${\bf I'}$ Symmetry and the Quark Sector}

The irreps of $I'$  are one singlet $\textbf{1}$, two triplets $\textbf{3}$ and $\textbf{3}'$, one quartet $\textbf{4}$, and one quintet $\textbf{5}$, which are also the irreps of $A_{5}$, plus $I'$ has four spinor-like irreps $\textbf{2}_{s}, \textbf{2}'_{s}, \textbf{4}_{s}$, and $\textbf{6}_{s}$. The characters and the multiplication rules of $I'$ symmetry can be found in Table~\ref{character} and Table~\ref{multiplication} of the Appendix.   

In this work we confine ourselves to minimally extending the $A_{5}$ model, i.e., to include the four generations of quarks while minimizing the introduction of other new degrees of freedom. The assignment of the quark sector under $I'\times Z_{2}\times Z_{3}$ is given as follows:  
\begin{eqnarray}
\underbrace{\left(\begin{array}{c} u \\ d \end{array}\right)_{L} 
\left(\begin{array}{c} c \\ s \end{array}\right)_{L}}_{U_{1L}(\textbf{2}_{s},+1,\omega)} \quad {\rm and} \quad
\underbrace{\left(\begin{array}{c} t \\  b \end{array}\right)_{L}\left(\begin{array}{c} t' \\  b' \end{array}\right)_{L}}_{U_{2L}(\textbf{2}_{s},+1,\omega^2)}
\end{eqnarray}
for the left-handed doublets, and   
\begin{eqnarray}
\underbrace{\underbrace{d_{R}, s_{R}}_{S_{R}}\underbrace{u_{R}, c_{R}}_{C_{R}}}_{D_{sR}(\textbf{4}_{s},+1,+1)} \quad {\rm ,} \quad \underbrace{b_{R}, b'_{R}}_{D_{bR}(\textbf{2}'_{s},-1,\omega^2)} \quad {\rm ,} \quad \underbrace{t_{R}, t'_{R}}_{D_{tR}(\textbf{2}'_{s},+1,\omega^2)} 
\end{eqnarray}
for the right-handed singlets. Here the fields $(t' , b')^{T}_{L}$, $b'_{R}$ and $t'_{R}$ denote the chiral fields of the fourth generation quarks and the Higgs sector is the same as the $A_{5}$ model. Thus, we can write  the most general Yukawa interactions between quarks and scalar fields as 
%
\begin{eqnarray}\label{YukawaQ}
L_{\rm Y(quark)} & = & f_{1}(U_{1L}\otimes D_{sR})\otimes\Phi_{3} + f_{2}(U_{2L}\otimes D_{bR})\otimes H'_{4}  \nonumber \\
&+& f_{3}(U_{2L}\otimes D_{tR})\otimes H_{4} + {\rm H.c.}
\end{eqnarray}
%
%

Recall that in the $A_{5}$ model the first step of symmetry breaking  $A_{5}\rightarrow A_{4}$ was caused by the vacuum expectation values (VEVs) of the $SU(2)_{L}$ singlet scalars $\langle S_{4}\rangle = (V_{S}, 0, 0, 0)$. Here the $S_4$ VEV breaks $I'$ to $T'$ symmetry.   
The decomposition of the irreps at this stage of symmetry breaking $I'\rightarrow T'$ is given in Table \ref{sb} in the Appendix.
Therefore, the quark fields decompose as
\begin{eqnarray}
U_{1L}(\textbf{2}_{s},+1,\omega) & \rightarrow & U_{1L}(\textbf{2}, +1, \omega) 
\;\;  ,\nonumber\\
U_{2L}(\textbf{2}_{s}, +1, \omega^2) & \rightarrow & U_{2L}(\textbf{2}, +1, \omega^2)
\; \;  , \nonumber \\
D_{sR}(\textbf{4}_{s}, +1, +1) & \rightarrow & S_{R}(\textbf{2}', +1, +1) + C_{R}(\textbf{2}'', +1, +1) \; \; , \nonumber \\
D_{bR}(\textbf{2}'_{s}, -1, \omega^2) & \rightarrow & D_{bR}(\textbf{2}, -1, \omega^2) 
\;\; , \nonumber \\
D_{tR}(\textbf{2}'_{s}, +1, \omega^2) & \rightarrow & D_{tR}(\textbf{2}, +1, \omega^2) \;\; ; 
\end{eqnarray}
while for the scalars we have 
\begin{eqnarray}
\label{A4scalar}
S_{4}(\textbf{4},+1,+1) & \rightarrow & S_{1}(\textbf{1},+1,+1) + S_{3}(\textbf{3}, +1, +1) \; \; , \nonumber \\
H_{4}(\textbf{4}, +1, \omega^2) & \rightarrow & H_{1}(\textbf{1}, +1, \omega^2) + H_{3}(\textbf{3}, +1, \omega^2) \; \; , \nonumber \\
H'_{4}(\textbf{4}, -1, \omega^2) & \rightarrow & H'_{1}(\textbf{1}, -1, \omega^2) + H'_{3}(\textbf{3}, -1, \omega^2) \; \; , \nonumber \\
\Phi_{3}(\textbf{3}, +1, \omega^2) & \rightarrow &\Phi_{3}(\textbf{3}, +1, \omega^2) \;\; . 
\end{eqnarray}

The Yukawa interactions of Eq.~(\ref{YukawaQ}) now reads  
\begin{eqnarray}
f_{1}[U_{1L(\textbf{2})} & \otimes & (S_{R(\textbf{2}')} \oplus C_{R(\textbf{2}'')})]\otimes\Phi_{3} 
\nonumber \\
+ f_{2}[U_{2L(\textbf{2})}& \otimes & D_{bR(\textbf{2})}]\otimes(H'_{1} \oplus H'_{3})  \\
+ f_{3}[U_{2L(\textbf{2})}& \otimes & D_{tR(\textbf{2})}]\otimes(H_{1}\oplus H_{3}) + {\rm H.c.}
\nonumber 
\end{eqnarray}
%
The Clebsch-Gordan coefficients of $T'$ can be found in the review~\cite{Ishimori:2010au} and were already calculated in~\cite{FFK,Frampton:2007et,Aranda:1999kc}. We adopt the coefficients shown in Ref.~\cite{Ishimori:2010au} and ignore the phases for simplicity. The Yukawa couplings are divided into terms involving the up-type quarks
%
\onecolumngrid
\begin{eqnarray}
&&f_{1}\left[-\left(\begin{array}{c}u \\d\end{array}\right)_{L}u_{R}
\Phi_{3_1} + \left(\begin{array}{c}c \\s\end{array}\right)_{L}c_{R}\Phi_{3_2} + \frac{1}{\sqrt{2}}
\big(\left(\begin{array}{c}u \\d\end{array}\right)_{L}c_{R} 
+ \left(\begin{array}{c}c \\s\end{array}\right)_{L}u_{R}\big)\Phi_{3_3}\right] 
\nonumber \\
&+& \frac{f_{3}}{\sqrt{2}}\left[\left(\begin{array}{c}t \\b\end{array}\right)_{L}t'_{R}
\big(H_{1} + H_{3_1}\big) - \left(\begin{array}{c}t' \\b'\end{array}\right)_{L}t_{R}
\big(H_{1} - H_{3_1}\big)\right]  \\
&+& f_{3}\left[-\left(\begin{array}{c}t \\b\end{array}\right)_{L}t_{R}H_{3_2} 
+ \left(\begin{array}{c}t' \\b'\end{array}\right)_{L}t'_{R}H_{3_3} \right] \nonumber
\end{eqnarray}
and the down-type quarks 
\onecolumngrid
\begin{eqnarray}
&&f_{1}\left[\left(\begin{array}{c}c \\s\end{array}\right)_{L}s_{R}\Phi_{3_1} + \frac{1}{\sqrt{2}}\big(\left(\begin{array}{c}u \\d\end{array}\right)_{L}s_{R} 
 + \left(\begin{array}{c}c \\s\end{array}\right)_{L}d_{R}\big)
\Phi_{3_2} - \left(\begin{array}{c}u \\d\end{array}\right)_{L}d_{R}\Phi_{3_3}
\right] \nonumber \\
&+& \frac{f_{2}}{\sqrt{2}}\left[\left(\begin{array}{c}t \\b\end{array}\right)_{L}b'_{R}\big(H'_{1} + H'_{3_1}\big) - \left(\begin{array}{c}t' \\b'\end{array}\right)_{L}b_{R}\big(H'_{1} - H'_{3_1}\big)\right]  \\
&+& f_{2}\left[-\left(\begin{array}{c}t \\b\end{array}\right)_{L}b_{R}H'_{3_2} + \left(\begin{array}{c}t' \\b'\end{array}\right)_{L}b'_{R}H'_{3_3}\right] \nonumber
\end{eqnarray}
%
\noindent
respectively. Here we express the components of scalar fields as 
$\Phi_{3} = (\Phi_{3_1}, \Phi_{3_2}, \Phi_{3_3})$, 
$H_{4} = H_{1} + H_{3}$ where $H_{3} = (H_{3_1}, H_{3_2}, H_{3_3})$, and 
$H'_{4} = H'_{1} + H'_{3}$ where $H'_{3}= (H'_{3_1}, H'_{3_2}, H'_{3_3})$ 
according to the breaking $I' \rightarrow T'$ shown in Eq.~(\ref{A4scalar}). 
The masses of up-type fermions are generated by the $\Phi_3$ and $H_{4}$ VEVs, 
while down-type fermion masses are generated by the $\Phi_3$ and $H'_{4}$ VEVs. 

\section{Masses and Mixings}  

We notice that each up- and down-type quark mass matrix is divided into two $2\times2$ 
block matrices and can be expressed as 
\begin{eqnarray}
M_{U} = \left(\begin{array}{c|c}M_{uc} & 0 \\\hline 0 & M_{tt'}\end{array}\right) \quad {\rm and} \quad M_{D} = \left(\begin{array}{c|c}M_{ds} & 0 \\\hline 0 & M_{bb'}\end{array}\right).
\end{eqnarray}
This indicates the first two generations mix only with the third and fourth generations through  higher order corrections. The $2\times2$ block mass matrices are given by  
\begin{eqnarray}
M_{uc} = f_{1}\left(\begin{array}{cc}-\langle \Phi_{3_1}\rangle & \frac{\langle \Phi_{3_3}\rangle}{\sqrt{2}} \\\frac{\langle \Phi_{3_3}\rangle}{\sqrt{2}} & \langle\Phi_{3_2}\rangle\end{array}\right)
\;\; , \; \;  M_{ds} = f_{1}\left(\begin{array}{cc}-\langle\Phi_{3_3}\rangle & \frac{\langle \Phi_{3_2}\rangle}{\sqrt{2}} \\\frac{\langle \Phi_{3_2}\rangle}{\sqrt{2}} & \langle\Phi_{3_1}\rangle\end{array}\right), 
\end{eqnarray}
\begin{eqnarray}
M_{tt'} = f_{3}\left(\begin{array}{cc}-\langle H_{3_2}\rangle & \frac{1}{\sqrt{2}}(\langle H_{1}\rangle + \langle H_{3_1}\rangle) \\\frac{1}{\sqrt{2}}(-\langle H_{1}\rangle + \langle H_{3_1}\rangle) & \langle H_{3_3}\rangle\end{array}\right),
\end{eqnarray}
and 
\begin{eqnarray}
M_{bb'} = f_{2}\left(\begin{array}{cc}-\langle H'_{3_2}\rangle & \frac{1}{\sqrt{2}}(\langle H'_{1}\rangle + \langle H'_{3_1}\rangle) \\\frac{1}{\sqrt{2}}(-\langle H'_{1}\rangle + \langle H'_{3_1}\rangle) & \langle H'_{3_3}\rangle\end{array}\right).
\end{eqnarray}
%
With the  $\langle\Phi_{3}\rangle = v(1,1,1)$ VEV, which is enforced by the requirement of tribimaximal mixings in neutrinos at leading order~\cite{Chen:2010ty}, we find that the masses of $u, d, s, c$ are degenerate $m_{u} = m_{d} = m_{s} = m_{c} = \sqrt{\frac{3}{2}}f_{1}v$.
Also, the mass matrices of both $M_{uc}$ and $M_{ds}$ take the same form. Therefore, the matrices $M_{uc}M^{\dag}_{uc}$ and $M_{ds}M^{\dag}_{ds}$ are diagonalized by using the same unitary matrix. Thus we conclude that, at first order, the CKM matrix is forced to be the identity\footnote{The third and fourth generations can mix largely in principle.}, which is an acceptable first approximation to $V_{\rm CKM}$. It is interesting  that the light quark masses $(u,d,s,c)$ and three SM light neutrino masses\footnote{Neutrino masses are generated through the seesaw mechanism, and therefore they are further suppressed by the lepton number breaking scale.} have the same origin; both come from the VEVs of $\Phi_{3}$ field, and the unit CKM matrix and the tribimaximal PMNS mixings arise from the single subgroup $Z_{3}$, which is the remnant symmetry left in the vacuum $\langle\Phi_{3}\rangle = v(1,1,1)$.
To correct the CKM mixings  by the high-order effects, the relevant dimension-five and -six effective operators are 
$U_{2L}D_{sR}\Phi^2_{3}$, $U_{2L}D_{sR}\Phi_{3}H_{4}$, 
$U_{2L}D_{sR}H'^2_{4}$, $U_{2L}D_{sR}H^2_{4}$
and
$U_{1L}D_{bR}H^2_{4}H'_{4}$, $U_{1L}D_{tR}H^3_{4}$,
$U_{1L}D_{tR}H'^2_{4}H_{4}$ 
respectively. Also, as mentioned in Ref.~\cite{Chen:2010ty},  perturbations of the $\Phi_{3}$ VEVs   are needed to accommodate realistic neutrino masses. {These high-order corrections will link together
the derivations of the Cabibbo angle $\theta_{C}$ and $\theta_{13}$
in the quark and lepton mixing matrices respectively.  A nonzero value of  $\theta_{13}$ has recently been indicated by several experiments~\cite{Abe:2011sj,Abe:2011fz,An:2012eh,Ahn:2012nd} and by global analyses ~\cite{Fogli:2011qn}}. If we consider perturbations of the VEVs  by taking $\langle \Phi_{3} \rangle = (v + \Delta_{1}, v + \Delta_{2}, v)$, the light quark masses are calculated to be
\begin{eqnarray}
m^2_{u,c} & = &\frac{f^2_1}{2}\bigg[ 3v^2 + 2v(\Delta_{2} - \Delta_{1}) + (\Delta^2_{1} + \Delta^2_{2}) \nonumber \\
&\mp &\sqrt{(\Delta^2_{1} + \Delta^2_{2})^2(6v^2 - 4v(\Delta_{1} + \Delta_{2}))}\bigg]
\nonumber
\end{eqnarray}  
and
\begin{eqnarray}
m^2_{d,s} & = &\frac{f^2_{1}}{2}\bigg[ 3v^2 + 2v(\Delta_{1} + \Delta_{2}) + (\Delta^2_{1} + \Delta^2_{2}) \nonumber \\
& \mp & \Delta_{1}\sqrt{(2v + \Delta_{1})^2 + 2(v + \Delta_{2})^2}\bigg] \nonumber \; \; .
\end{eqnarray}
This indicates how the degeneracy of the  light quark masses can be lifted.

Recall that in the lepton $A_5$ model, $H_{4}$ is responsible for the Dirac masses of neutrinos, and we require the condition $\langle H_{1}\rangle \equiv V_{1} \gg \langle H_{3_{1,2,3}}\rangle \equiv V_{3_{1,2,3}}$ in order to decouple the 4th generation neutrino from the three light SM neutrinos. Therefore,
from $M_{t,t'}$, we obtain the masses of $t$ and $t'$ as 
$m^2_{t,t'} \approx \left[ V^2_{1} + (V^2_{3_1} + V^2_{3_2} + V^2_{3_3}) \mp V_{1}
\sqrt{4V^2_{3_1} + 2(V_{3_2} + V_{3_3})^2} \, \right]/2$. 
For $M_{bb'}$, we also follow the lepton $A_5$ model by taking~\footnote{The four component VEVs, in general, can be different.}
$\langle H'_{4}\rangle = (V'_{1}, V',V',V')$, since this gives masses to charged leptons too.
$m^2_{b,b'}$ are then given by 
$[{V'^2_{1} + 3V'^2 \mp 2\sqrt{3}V'_{1}V'}]/{2}$. In general, we have enough parameters to accommodate the heavy quark mass spectrum. 

\section{Conclusion} 

As mentioned in the introduction, the recent observations of a boson with a mass near 125 GeV \cite{atlas,cms} have placed severe constraints
on the standard model augmented by a sequential fourth generation of fermions. 
The CMS experiment has now excluded such a fourth generation of fermions
with masses of up to 600 GeV \cite{cms-search}.
Note that our $I'$ model has three doublets and one singlet Higgs, all of which are in non-singlet irreps of $I'$. Hence it is not necessarily disfavored by the current experimental search.
Indeed, such severe limits can be relaxed into the range of $400 \sim 600$ GeV in a two Higgs doublets model with four generation of fermions, as discussed in Ref.\cite{BarShalom:2012ms}. 
It would be interesting to investigate whether our $I'$ model (or one of its extensions) can be recast into a form designed in Ref.\cite{BarShalom:2012ms}.

In summary, we construct a model of four fermion generations based on the 
binary icosahedral symmetry group $I'$. Many properties of the SM with three families are accommodated 
such as the mass spectrum, tribimaximal mixings in the neutrino sector, and an identity CKM matrix at leading order\footnote{Both $A_5$ and $I'$ have also recently been used in the context of three generation
models of fermion masses and mixings~\cite{Ding:2011cm}.}. In addition, quarks and leptons relations are intimately connected as their masses are provided from the same set of scalars. We believe this makes the model both interesting and challenging. For example, one has to strike a balance between the result of tribimaximal mixings in the neutrino sector and derivation a
realistic Cabibbo angle from perturbations in the quark sector. It is still not clear to us  whether the higher order corrections will lead to a realistic Cabibbo angle or if we need extra degrees of freedom to realize it. 
We will leave it and other phenomenological aspects of this model to future work.


\vskip 0.5in

\acknowledgments
This work was supported in part by the US DOE grant DE-FG05-85ER40226, the
National Science Council of Taiwan under Grant Numbers 98-2112-M-001-014-MY3,
101-2112-M-001-005-MY3, and the National Center for Theoretical Sciences of Taiwan (NCTS). 
TWK is grateful for the hospitality of the Physics Division of NCTS where this work was initiated.
TCY thanks the hospitalities of KITPC (Beijing, China) and IMSc (Chennai, India) 
where progress of this work was made.


\section{Appendix}

\subsection*{Discrete Symmetry Groups $I'$ and $A_{5}$}
$A_{5}$ is the only simple finite discrete subgroup of $SO(3)$. Its $60$ elements can be generated by products of the two generators $a$ and $b$, which satisfy 
\begin{eqnarray}
a^2 = b^3 = (ab)^5 = e \; ,
\end{eqnarray}
where $e$ is the identity element.

\begin{figure}[t]
  \centering
    \includegraphics[width=0.5\textwidth]{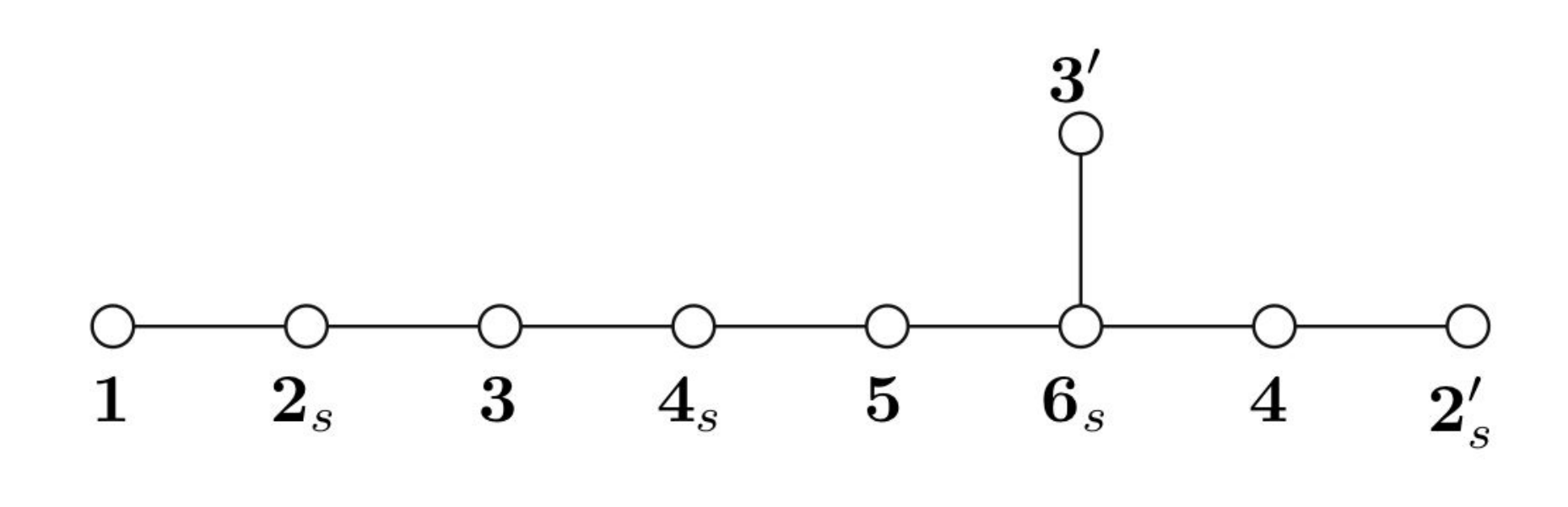}
  \caption{Extended Dynkin diagram and irreducible representations of $I'$.}
  \label{fig:dynkin}
\end{figure}

$I'$ is the double covering of $A_{5}$, therefore it has 120 elements. The representation in terms of generators is similar to that of $A_{5}$, namely
\begin{eqnarray}
 a^2 = b^3 = (ab)^5 \; .
\end{eqnarray}
We note that $a^2$ is no longer the identity, but the negative of the identity, i.e., $a^4=b^6=e$. Any finite subgroup of $SU(2)$ must have (at least) one spinor doublet $\textbf{2}_{s}$. By using the multiplication rules, the irreducible representations of the group~\cite{Luhn:2007yr, p.ramond} can be visualized as 
an extended Dynkin diagrams depicted in Fig.~\ref{fig:dynkin}.

Recalling that the exact sequence relation between $SU(2)$ and $SO(3)$ is
\begin{eqnarray}
1\rightarrow Z_2 \rightarrow SU(2) \rightarrow SO(3) \rightarrow 1 \; \; ,
\end{eqnarray}
we can restrict to the discrete cases
\begin{eqnarray}
1\rightarrow Z_2 \rightarrow T' \rightarrow T \rightarrow 1
\end{eqnarray}
and
\begin{eqnarray}
1\rightarrow Z_2 \rightarrow I' \rightarrow I \rightarrow 1
\end{eqnarray}
to demonstrate the double coverings. As our interest is in $I'$ we first reproduce its character table in 
Table \ref{character}, from which we can easily calculate the  multiplication table for $I'$ irreps 
presented in Table \ref{multiplication}. 
The symmetry breaking to $T'$ is also easily obtained, as given in Table \ref{sb}.

\onecolumngrid
\begin{center}
\begin{table}[h]
\begin{tabular}{|c|c|c|c|c|c|c|c|c|c|}\hline
 & $C_{1}(1)$ & $C_{2}(1)$ & $C_{3}(12)$ & $C_{4}(12)$ & $C_{5}(12)$ & $C_{6}(12)$ & $C_{7}(30)$ & $C_{8}(20)$ & $C_{9}(20)$ \\ \hline
\textbf{1} & $1$ & $1$ & $1$ & $1$ & $1$ & $1$ & $1$ & $1$ & $1$  \\ 
\textbf{3} & $3$ & $3$ & $1 - \phi$ & $1 - \phi$ & $\phi$ & $\phi$ & $-1$ & $0$ & $0$ \\
$\textbf{3}'$ & $3$ & $3$ & $\phi$ & $\phi$ & $1 - \phi$ & $1 - \phi$ & $-1$ & $0$ & $0$ \\ 
\textbf{4} & $4$ & $4$ & $-1$ & $-1$ & $-1$ & $-1$ & $0$ & $1$ & $1$ \\ 
\textbf{5} & $5$ & $5$ & $0$ & $0$ & $0$ & $0$ & $1$ & $-1$ & $-1$  \\\hline
$\textbf{2}_{s}$ & $2$ & $-2$ & $\phi -1$ & $1 - \phi$ & $-\phi$ & $\phi$ & $0$ & $-1$ & $1$ \\ 
$\textbf{2}_{s}'$ & $2$ & $-2$ & $-\phi$ & $\phi$ & $\phi -1$ & $1 - \phi$ & $0$ & $-1$ & $1$ \\
$\textbf{4}_{s}$ & $4$ & $-4$ & $-1$ & $1$ & $-1$ & $1$ & $0$ & $1$ & $-1$ \\
$\textbf{6}_{s}$ & $6$ & $-6$ & $1$ & $-1$ & $1$ & $-1$ & $0$ & $0$ & $0$\\ \hline\hline
\end{tabular}
\caption{\label{character} Character table of $I'$ where $\phi = \frac{1 + \sqrt{5}}{2}$ 
is the golden ratio.}
\end{table}
\end{center} 


\begin{center}
\begin{table}[h]
\begin{tabular}{|c||c|p{1.8cm}|p{1.8cm}|p{1.8cm}|p{1.8cm}||p{1.8cm}|p{1.8cm}|p{1.8cm}|p{1.8cm}|}\hline 
 $\otimes$& \textbf{1} & \textbf{3} & $\textbf{3}'$ & \textbf{4} & \textbf{5} & $\textbf{2}_{s}$ & $\textbf{2}'_{s}$ & $\textbf{4}_{s}$ & $\textbf{6}_{s}$ \\ \hline \hline 
 \textbf{1}& \textbf{1} & \textbf{3} & $\textbf{3}'$ & \textbf{4} & \textbf{5} & $\textbf{2}_{s}$ & $\textbf{2}'_{s}$ & $\textbf{4}_{s}$ & $\textbf{6}_{s}$  \\ \hline 
\textbf{3} & \textbf{3} & $\textbf{1}\oplus\textbf{3}\oplus\textbf{5}$  & $\textbf{4}\oplus\textbf{5}$ & $\textbf{3}'\oplus\textbf{4}\oplus\textbf{5}$ & $\textbf{3}\oplus\textbf{3}'\oplus\textbf{4}\oplus\textbf{5}$ & $\textbf{2}_{s}\oplus\textbf{4}_{s}$ & $\textbf{6}_{s}$ & $\textbf{2}_{s}\oplus\textbf{4}_{s}\oplus\textbf{6}_{s}$ & $\textbf{2}'_{s}\oplus\textbf{4}_{s}\oplus\textbf{6}_{s}\oplus\textbf{6}_{s}$ \\ \hline 
$\textbf{3}'$ & $\textbf{3}'$ & $\textbf{4}\oplus\textbf{5}$ & $\textbf{1}\oplus\textbf{3}'\oplus\textbf{5}$ & $\textbf{3}\oplus\textbf{4}\oplus\textbf{5}$ & $\textbf{3}\oplus\textbf{3}'\oplus\textbf{4}\oplus\textbf{5}$ & $\textbf{6}_{s}$ & $\textbf{2}'_{s}\oplus\textbf{4}_{s}$ & $\textbf{2}'_{s}\oplus\textbf{4}_{s}\oplus\textbf{6}_{s}$ & $\textbf{2}_{s}\oplus\textbf{2}'_{s}\oplus\textbf{4}_{s}\oplus\textbf{6}_{s}$ \\ \hline 
\textbf{4} & \textbf{4} & $\textbf{3}'\oplus\textbf{4}\oplus\textbf{5}$ & $\textbf{3}\oplus\textbf{4}\oplus\textbf{5}$ & $\textbf{1}\oplus\textbf{3}\oplus\textbf{3}'\oplus\textbf{4}\oplus\textbf{5}$ & $\textbf{3}\oplus\textbf{3}'\oplus\textbf{4}\oplus\textbf{5}\oplus\textbf{5}$ & $\textbf{2}'_{s}\oplus\textbf{6}_{s}$ & $\textbf{2}_{s}\oplus\textbf{6}_{s}$ & $\textbf{4}_{s}\oplus\textbf{6}_{s}\oplus\textbf{6}_{s}$ & $\textbf{2}_{s}\oplus\textbf{2}'_{s}\oplus\textbf{4}_{s}\oplus\textbf{4}_{s}\oplus\textbf{6}_{s}\oplus\textbf{6}_{s}$ \\ \hline
\textbf{5} & \textbf{5} & $\textbf{3}\oplus\textbf{3}'\oplus\textbf{4}\oplus\textbf{5}$ & $\textbf{3}\oplus\textbf{3}'\oplus\textbf{4}\oplus\textbf{5}$ & $\textbf{3}\oplus\textbf{3}'\oplus\textbf{4}\oplus\textbf{5}\oplus\textbf{5}$ & $\textbf{1}\oplus\textbf{3}\oplus\textbf{3}'\oplus\textbf{4}\oplus\textbf{4}\oplus\textbf{5}\oplus\textbf{5}$ & $\textbf{4}_{s}\oplus\textbf{6}_{s}$ & $\textbf{4}_{s}\oplus\textbf{6}_{s}$ & $\textbf{2}_{s}\oplus\textbf{2}'_{s}\oplus\textbf{4}_{s}\oplus\textbf{6}_{s}\oplus\textbf{6}_{s}$ & $\textbf{2}_{s}\oplus\textbf{2}'_{s}\oplus\textbf{4}_{s}\oplus\textbf{4}_{s}\oplus\textbf{6}_{s}\oplus\textbf{6}_{s}\oplus\textbf{6}_{s}$ \\\hline \hline
$\textbf{2}_{s}$ & $\textbf{2}_{s}$ & $\textbf{2}_{s}\oplus\textbf{4}_{s}$ & $\textbf{6}_{s}$ & $\textbf{2}'_{s}\oplus\textbf{6}_{s}$ & $\textbf{4}_{s}\oplus\textbf{6}_{s}$ & $\textbf{1}\oplus\textbf{3}$ & \textbf{4} & $\textbf{3}\oplus\textbf{5}$ & $\textbf{3}'\oplus\textbf{4}\oplus\textbf{5}$ \\ \hline
$\textbf{2}'_{s}$ & $\textbf{2}'_{s}$ & $\textbf{6}_{s}$ & $\textbf{2}'_{s}\oplus\textbf{4}_{s}$ & $\textbf{2}_{s}\oplus\textbf{6}_{s}$ & $\textbf{4}_{s}\oplus\textbf{6}_{s}$ & $\textbf{4}$ & $\textbf{1}\oplus\textbf{3}'$ & $\textbf{3}'\oplus\textbf{5}$ & $\textbf{3}\oplus\textbf{4}\oplus\textbf{5}$ \\ \hline
$\textbf{4}_{s}$ & $\textbf{4}_{s}$ & $\textbf{2}_{s}\oplus\textbf{4}_{s}\oplus\textbf{6}_{s}$ & $\textbf{2}'_{s}\oplus\textbf{4}_{s}\oplus\textbf{6}_{s}$ & $\textbf{4}_{s}\oplus\textbf{6}_{s}\oplus\textbf{6}_{s}$ & $\textbf{2}_{s}\oplus\textbf{2}'_{s}\oplus\textbf{4}_{s}\oplus\textbf{6}_{s}\oplus\textbf{6}_{s}$ & $\textbf{3}\oplus\textbf{5}$ & $\textbf{3}'\oplus\textbf{5}$ & $\textbf{3}'\oplus\textbf{4}\oplus\textbf{5}$ & $\textbf{3}\oplus\textbf{3}'\oplus\textbf{4}\oplus\textbf{4}\oplus\textbf{5}\oplus\textbf{5}$ \\ \hline 
$\textbf{6}_{s}$ & $\textbf{6}_{s}$ & $\textbf{2}'_{s}\oplus\textbf{4}_{s}\oplus\textbf{6}_{s}\oplus\textbf{6}_{s}$ & $\textbf{2}_{s}\oplus\textbf{2}'_{s}\oplus\textbf{4}_{s}\oplus\textbf{6}_{s}$ & $\textbf{2}_{s}\oplus\textbf{2}'_{s}\oplus\textbf{4}_{s}\oplus\textbf{4}_{s}\oplus\textbf{6}_{s}\oplus\textbf{6}_{s}$ & $\textbf{2}_{s}\oplus\textbf{2}'_{s}\oplus\textbf{4}_{s}\oplus\textbf{4}_{s}\oplus\textbf{6}_{s}\oplus\textbf{6}_{s}\oplus\textbf{6}_{s}$ & $\textbf{3}'\oplus\textbf{4}\oplus\textbf{5}$ & $\textbf{3}\oplus\textbf{4}\oplus\textbf{5}$ & $\textbf{3}\oplus\textbf{3}'\oplus\textbf{4}\oplus\textbf{4}\oplus\textbf{5}\oplus\textbf{5}$ & $\textbf{1}\oplus\textbf{3}\oplus\textbf{3}\oplus\textbf{3}'\oplus\textbf{3}'\oplus\textbf{4}\oplus\textbf{4}\oplus\textbf{5}\oplus\textbf{5}\oplus\textbf{5}$\\ \hline
\end{tabular}
\caption{\label{multiplication} Multiplication rules for the binary icosahedral group $I'$.}
\end{table}
\end{center} 


\begin{center}
\begin{table}[h]
\begin{tabular}{|p{1.0cm}cc|p{1.0cm}cc|}\hline
$I'$ & $\rightarrow$ & $T'$ & $I'$ & $\rightarrow$ & $T'$ \\ \hline
$\textbf{1}$ & & $\textbf{1} $ & $\textbf{2}_{s}$ & & $\textbf{2}$ \\ 
$\textbf{3}$ & & $\textbf{3} $& $\textbf{2}'_{s}$ & & $\textbf{2}$ \\ 
$\textbf{3}'$ & & $\textbf{3} $ & $\textbf{4}_{s}$ & & $\textbf{2}' + \textbf{2}''$ \\ 
$\textbf{4}$ & & $\textbf{1} + \textbf{3} $ & $\textbf{6}_{s}$ & & $\textbf{2} + \textbf{2}' + \textbf{2}''$ \\ 
$\textbf{5}$ & & $\textbf{1}' + \textbf{1}'' + \textbf{3} $ & & & \\\hline
\end{tabular}
\caption{\label{sb}  $I'\rightarrow T'$ symmetry breaking.}
\end{table}
\end{center} 

\subsection*{Higgs Potential}
The most general form of the Higgs potential containing the scalar fields $S_{4}$, $H_{4}$, $H'_{4}$ and $\Phi_{3}$, 
invariant under the discrete $A_{5}\times Z_{2}\times Z_{3}$ symmetries is given  by
\begin{eqnarray}
V &=& V(S_{4}) + V(H_{4}) + V(H'_{4}) + V(\Phi_{3}) + V(S_{4}, H_{4}) \nonumber \\ 
&& + V(S_{4}, H'_{4}) + V(S_{4}, \Phi_{3}) + V(H_{4}, H'_{4}) \nonumber \\ 
&& + V(H_{4}, \Phi_{3}) + V(H'_{4}, \Phi_{3}) + V(H_{4}, H'_{4}, \Phi_{3}) 
\end{eqnarray} 
where the individual terms are written as 
\begin{eqnarray*}
V(S_{4}) &=& \mu^2_{S_4}S^2_{4} + \mu_{s}(S^2_{4})_{\textbf{4}}S_{4} + \lambda^{s}_{\alpha}(S^2_{4})_{\alpha}(S^2_{4})_{\alpha}, \\
V(H_{4}) &=& \mu^2_{H}(H^{\dag}_{4}H_{4})_{\textbf{1}} + \lambda^{H}_{\alpha}(H^{\dag}_{4}H_{4})_{\textbf{$\alpha$}}(H^{\dag}_{4}H_{4})_{\textbf{$\alpha$}}, \\ 
V(H'_{4}) &=& \mu^2_{H'}(H'^{\dag}_{4}H'_{4})_{\textbf{1}} + \lambda^{H'}_{\alpha}(H'^{\dag}_{4}H'_{4})_{\textbf{$\alpha$}}(H'^{\dag}_{4}H'_{4})_{\textbf{$\alpha$}}, \\ 
V(\Phi_{3}) &=& \mu^2_{\Phi}(\Phi^{\dag}_{3}\Phi_{3})_{\textbf{1}} + \lambda^{\Phi}_{\beta}(\Phi^{\dag}_{3}\Phi_{3})_{\textbf{$\beta$}}(\Phi^{\dag}_{3}\Phi_{3})_{\textbf{$\beta$}}, \\ 
V(S_{4}, H_{4}) &=& \delta^{HS}(H^{\dag}_{4}H_{4})_{\textbf{4}}S_{4} + \lambda^{HS}_{\alpha}(H^{\dag}_{4}H_{4})_{\textbf{$\alpha$}}(S^2_4)_{\textbf{$\alpha$}}, \\ 
V(S_{4}, H'_{4}) &=& \delta^{H'S}(H'^{\dag}_{4}H'_{4})_{\textbf{4}}S_{4} + \lambda^{H'S}_{\alpha}(H'^{\dag}_{4}H'_{4})_{\textbf{$\alpha$}}(S^2_4)_{\textbf{$\alpha$}}, \\  
V(S_{4}, \Phi_{3}) &=& \delta^{\Phi S}(\Phi^{\dag}_{3}\Phi_{3})_{\textbf{4}}S_{4} + \lambda^{\Phi S}_{\beta}(\Phi^{\dag}_{3}\Phi_{3})_{\textbf{$\beta$}}(S^2_{4})_{\textbf{$\beta$}}, \\ 
V(H_{4}, H'_{4}) &=& \lambda^{HH'}_{\alpha}(H^{\dag}_{4}H_{4})_{\textbf{$\alpha$}}(H'^{\dag}_{4}H'_{4})_{\textbf{$\alpha$}} \nonumber \\ 
&& + \lambda'^{HH'}_{\alpha}(H^{\dag}_{4}H'_{4})_{\textbf{$\alpha$}}(H'^{\dag}_{4}H_{4})_{\textbf{$\alpha$}} \nonumber \\
&& + \left[ \lambda''^{HH'}_{\alpha}(H^{\dag}_{4}H'_{4})_{\textbf{$\alpha$}}(H^{\dag}_{4}H'_{4})_{\textbf{$\alpha$}} + \rm{H.c.}\right], \\
\end{eqnarray*} 
\begin{eqnarray*}
V(H_{4}, \Phi_{3}) &=& \lambda^{H\Phi}_{\beta}(H^{\dag}_{4}H_{4})_{\textbf{$\beta$}}(\Phi^{\dag}_{3}\Phi_{3})_{\textbf{$\beta$}} \nonumber \\
&& + \lambda'^{H\Phi}_{\gamma}(H^{\dag}_{4}\Phi_{3})_{\textbf{$\gamma$}}(\Phi^{\dag}_{3}H_{4})_{\textbf{$\gamma$}} \nonumber \\
&& + \left[ \lambda''^{H\Phi}_{\gamma}(H^{\dag}_{4}\Phi_{3})_{\textbf{$\gamma$}}(H^{\dag}_{4}\Phi_{3})_{\textbf{$\gamma$}} + \rm{H.c.} \right],  \\
V(H'_{4}, \Phi_{3}) &=& \lambda^{H'\Phi}_{\beta}(H'^{\dag}_{4}H'_{4})_{\textbf{$\beta$}}(\Phi^{\dag}_{3}\Phi_{3})_{\textbf{$\beta$}} \nonumber \\ 
&& + \lambda'^{H'\Phi}_{\gamma}(H'^{\dag}_{4}\Phi_{3})_{\textbf{$\gamma$}}(\Phi^{\dag}_{3}H'_{4})_{\textbf{$\gamma$}} \nonumber \\
&& + \left[ \lambda''^{H'\Phi}_{\gamma}(H'^{\dag}_{4}\Phi_{3})_{\textbf{$\gamma$}}(H'^{\dag}_{4}\Phi_{3})_{\textbf{$\gamma$}} + \rm{H.c.} \right],  \\
V(H_{4}, H'_{4}, \Phi_{3}) &=& \lambda^{HH'\Phi}_{\gamma}(H^{\dag}_{4}\Phi_{3})_{\textbf{$\gamma$}}(H'^{\dag}_{4}H'_{4})_{\textbf{$\gamma$}} \nonumber \\ 
&& + \lambda'^{HH'\Phi}_{\gamma}(H'^{\dag}_{4}\Phi_{3})_{\textbf{$\gamma$}}(H'^{\dag}_{4}H_{4})_{\textbf{$\gamma$}} \nonumber \\
&& + \lambda''^{HH'\Phi}_{\gamma}(H^{\dag}_{4}\Phi_{3})_{\textbf{$\gamma$}}(H^{\dag}_{4}H_{4})_{\textbf{$\gamma$}} + \rm{H.c..}
\end{eqnarray*}
Here we have introduced the $I'$ group representation indices $\alpha = \textbf{1}, \textbf{3}, \textbf{3}', \textbf{4}, \textbf{5}$; $\beta = \textbf{1}, \textbf{3}, \textbf{5}$; and $\gamma$ = $\textbf{3}', \textbf{4}, \textbf{5}$ respectively. The first stage of SSB takes us from the initial  $I'$ discrete symmetry to $T'$, and
this is accomplished with a VEV for $S_4$. Consider the pure $S_4$ sector of the Higgs
potential
$$
V_{S_4} = \mu_{S_4}^2 S_4S_4+ \lambda (S_4S_4)^2 + \lambda_1 (S_4 M_1S_4)^2+ \lambda_2 (S_4 M_3S_4)^2 .
$$
where $M_1$ and $M_3$ are $I'$ group matrices.
Since $(4\times 4)_S$ in $I'$ contains three terms there are three singlets in $[(4\times 4)_S]^2$ and hence in $V_{S_4}$.
There is also a potential cubic term that can be suppressed, either by imposing an addition $Z_2$
 symmetry, or by making $S_4$ a complex field. For $\lambda_1$ and $\lambda_2$ sufficiently small and of the proper signs, the SSB is dominated by the $\lambda$
 term and we have $\langle S_4 \rangle =-\sqrt{\frac{\mu_{S_4}^2}{\lambda }}(1,0,0,0)$ such that $I' \rightarrow T'$ with the SM gauge group unaffected and with the scalars decomposing as in Eq. (6).
 
 The sector of the Higgs potential of the $I'$ model that depends only on $H_1$, $H_{1'}$, $H_3$, and $H_{3'}$ is given in the appendix of \cite{Eby:2011qa} 
which only  involves $T'$ invariant terms and is identical to our form of the potential up to constraints due to residual $I'$ relations on the coupling constants.
The work of \cite{Eby:2011qa} demonstrates how $T'$ is broken completely and how the light quark and lepton masses and mixings arise.
 
 Our only other additional scalars are the $I'$ triplet, SM doublet fields $\Phi_3$. Since $3\rightarrow 3$ under $I' \rightarrow T'$, a $\Phi_3$ breaks $T'$.
 The pure $\Phi_3$ scalar sector can be rewritten as
 $$
 V_{\Phi_3} = \mu_{\Phi_3}^2 \Phi^{\dagger}_3\Phi_3+ \lambda (\Phi^{a\dagger}_3\Phi^a_3)^2 + \lambda' (\Phi^{a\dagger}_3\Phi^b_3)^2
 $$
 where we have included the triplet indices $a,b,...=1,2,3$ and find two quartic terms due to the fact that $(3\times 3)_S$ in $I'$ contains two terms and hence  $[(3\times 3)_S]^2$ contains two singlets. Given the $T'$ level VEVs for $H_3$, and $H_{3'}$, there is no $T'$ symmetry remaining to rotate the $\Phi_3$ VEV. Hence the choice $\langle \Phi_3 \rangle =-v (1,1,1)$ is stable for $\lambda' \approx 0$ and of the form needed in the model. 
 
A similar analysis can be applied to the investigation of the SSB in our $A_5$ model. Finally we note that typically, only an $O(10^{-1})$ fine tuning of scalar quartic coupling constants is needed to maintain the stability of such patterns of SSB.

\end{document}